\documentstyle[12pt]{article}
\def\be {\begin{equation}}
\def\ee {\end{equation}}
\def\ba {\begin{eqnarray}}
\def\ea {\end{eqnarray}}

\def\bi {\begin{itemize}}
\def\ei {\end{itemize}}
\begin{document}
\def\bea{\begin{eqnarray}}
\def\eea{\end{eqnarray}}
\title{\bf {Area Spectrum of Extremal Reissner-Nordstr\"om Black
Holes from Quasi-normal Modes }}
 \author{M.R. Setare  \footnote{E-mail: rezakord@ipm.ir}
  \\{Physics Dept. Inst. for Studies in Theo. Physics and
Mathematics(IPM)}\\
{P. O. Box 19395-5531, Tehran, IRAN }\\
 }

\maketitle
\begin{abstract}
Using the quasi-normal modes frequency of extremal
Reissner-Nordstr\"om black holes, we obtain area spectrum for
these type of black holes. We show that the area and entropy black
hole horizon are equally spaced. Our results for the spacing of
the area spectrum differ from that of schwarzschild black holes.
 \end{abstract}
\newpage

 \section{Introduction}
 The quantization of the black hole horizon area is one of the most interesting manifestations
 of quantum gravity. Since its first prediction by
Bekenstein  in 1974 \cite{bek1}, there has been much work on this
topic \cite{bek2}-\cite{bir}. Recently, the quantization of the
black hole area has been considered \cite{hod}, \cite{dry} as a
result of the absorption of a quasi-normal mode excitation. The
quasi-normal modes of black holes are the characteristic, ringing
frequencies which result from their perturbations \cite{chandra}
and provide a unique signature of these objects \cite{kokkotas},
possible to be observed in gravitational waves. In asymptotically
flat spacetimes the idea of QNMs started with the work of Regge
and Wheeler \cite{reggeW} where the stability of a black hole was
tested, and were first numerically computed by Chandrasekhar and
Detweiler several years later \cite{Chandra1}. The quasi-normal
modes bring now a lot of interest in different contexts: in
AdS/CFT correspondence \cite{Horowitz-Habeny}-\cite{Moss-Norman},
when considering thermodynamic properties of black holes in loop
quantum gravity \cite{dry}-\cite{motl}, in the context of possible
connection with
critical collapse \cite{Horowitz-Habeny,BHCC,kim}.\\
 Bekenstein's idea
for quantizing a black hole is based on the fact that its horizon
area, in the nonextreme case, behaves as a classical adiabatic
invariant \cite{bek1}, \cite{bekenstescola}. In the spirit of
Ehrenfest principle, any classical adiabatic invariant corresponds
to a quantum entity with discrete spectrum, Bekenstein conjectured
that the horizon area of a non extremal quantum black hole should
have a discrete eigenvalue spectrum. Moreover, the possibility of
a connection between the quasinormal frequencies of black holes
and the quantum properties of the entropy spectrum was first
observed by Bekenstein \cite{bek3}, and further developed by Hod
\cite{hod}. In particular, Hod proposed that the real part of the
quasinormal frequencies, in the infinite damping limit, might be
related via the correspondence principle to the fundamental quanta
of mass and angular momentum. The proposed correspondence between
quasinormal frequencies and the fundamental quantum of mass
automatically leads to an equally spaced area spectrum.
Remarkably, the spacing was such as to allow a statistical
mechanical interpretation for the resulting eigenvalues for the
Bekenstein-Hawking entropy. Dreyer\cite{dry} also used the large
damping quasi-normal mode frequency to fix the value of the
Immirzi parameter, $\gamma$, in loop quantum gravity. He found that loop quantum
gravity gives a correct prediction for the Bekenstein-Hawking entropy if gauge group
should be SO(3), and not SU(2).\\
In this letter our aim is to obtain the area and entropy spectrum
of extremal Reissner-Nordstr\"om  (RN) black holes in four
dimensional spacetime. Using the results of \cite{{nitz},{and}}
for highly damped quasi-normal modes we show how the horizon area
and entropy would be quantized. The authors of \cite{and} noted,
the variation of the mass of a RN black hole is not enough to
determine the variation of its area, since the corresponding
variation of the charge must be known. These authors, then,
assumed the same area quantum  as in the Schwarzshild case and
deduced the corresponding quantum of charge emission. It would
seem that in the case of an extremeal RN black hole this issue
does not arise, since $M$ and $Q$ are equal.  We show that the
results for the spacing of the area spectrum differ from
schwarzschild black hole case. Conversely, if we assume that
$\Delta A$ is indeed universal \cite{and} and thus remains as in
the Schwarzschild case $\Delta A=4\hbar \ln3$, then the real part
of the quasinormal frequency for extremal RN black hole is
different from  schwarzschild black hole case.

\section{Extremal Reissner-Nordstr\"om Black Holes}
The RN black hole's (event and inner) horizons are given in terms
of the black hole parameters by \be
r_{\pm}=M\pm\sqrt{M^{2}-Q^{2}}, \label{hor} \ee where $M$ and  $Q$
are respectively mass and charge of black hole. In the extreme
case these two horizons are coincides \be r_{\pm}=M, \hspace{1cm}
M=Q. \label{ext} \ee According a very interesting conclusion
follows \cite{nitz}(see also more recent paper \cite{and})  , the
real part of the quasinormal frequency for extremal RN black holes
coincides with the Schwarzschild value \be
\omega_{R}^{RN}=\frac{\ln 3}{4\pi R_{H}}, \label{quaf} \ee where
\be R_{H}=2M. \label{rfun} \ee We assume that this classical
frequency plays an important role in the dynamics of the black
hole and is relevant to its quantum properties \cite{{hod},{dry}}.
In particular, we consider $\omega_{R}^{RN}$ to be a fundamental
vibrational frequency for a black hole of energy $E=M$. Given a
system with energy $E$ and vibrational frequency $\omega$ one can
show that the quantity \be I=\int {dE\over \omega(E)},
\label{bohr} \ee is an adiabatic invariant \cite{kun},  which via
Bohr-Sommerfeld quantization has an equally spaced spectrum in the
semi-classical (large $n$) limit: \be I \approx n\hbar.
\label{smi} \ee Now by taking $\omega_{R}^{RN}$ in this context,
we have \be I=\int \frac{dE}{\omega_{R}^{RN}}=\int \frac{4\pi
R_{H}}{\ln 3}dM =\frac{4\pi}{\ln 3}\int 2M dM= \frac{4\pi}{\ln 3}
M^{2}+c, \label{adin} \ee where $c$ is a constant. In the other
hand, the black hole horizon area is given by \be A=4\pi
r_{+}^{2}. \label{area1} \ee Which using Eq.(\ref{ext}) in
extremal case is as following \be A=4\pi M^{2}. \label{area2} \ee
The Boher-Sommerfeld quantization law and Eq.(\ref{adin}) then
implies that the area spectrum is equally spaced, \be A_{n}=n
\hbar \ln 3.\label{spec} \ee By another method we can obtain above
result. From Eq.(\ref{area2} we get \be \Delta A =8\pi M \Delta
M=8\pi M \hbar  \omega_{R}^{RN} \label{delta}\ee where we have
associated the energy spacing with a frequency through $\Delta M
=\Delta E=\hbar\omega_{R}^{RN}$. Now using Eqs.(\ref{quaf},
\ref{rfun})we have \be \Delta A=\hbar \ln3, \label{spacing} \ee
therefore the extremal RN black hole have a discrete spectrum as
\be A_{n}=n\hbar \ln3. \label{spec2} \ee Which is exactly the
result of Eq.(\ref{spec}). Using the definition of the
Bekenstein-Hawking entropy we have \be
S=\frac{A_{n}}{4\hbar}=\frac{n\ln3}{4}. \label{entro} \ee The
above results for area spectrum and entropy is contradicted  by
results of Andersson and Howls \cite{and} for the extremal RN
black holes. Andersson and Howls have assumed that $\Delta A$ is
universal and thus remains as in the Schwarzschild case $\Delta
A=4\hbar \ln3$.\\
Now if we assume that $\Delta A$ is indeed universal \cite{and}
and thus remains as in the Schwarzschild case $\Delta A=4\hbar
\ln3$, then the real part of the quasinormal frequency for
extremal RN black hole is as \be \omega_{R}^{RN}=\frac{\ln 3}{\pi
R_{H}}, \label{quaf1} \ee in this case we have \be I=\int
\frac{dE}{\omega_{R}^{RN}}=\int \frac{\pi R_{H}}{\ln 3}dM
=\frac{\pi}{\ln 3}\int 2M dM= \frac{\pi}{\ln 3} M^{2}+c,
\label{adin1} \ee now  Eqs.(\ref{bohr}, \ref{area1},
\ref{area2},\ref{adin1})then implies that the area spectrum is
equally spaced as following, \be A_{n}=4n \hbar \ln
3.\label{spec1} \ee

  \section{Conclusion}
  Bekenestein's  idea for quantizing a black hole is based on the fact
  that its horizon area, in the nonextremal case,
behaves as a classical adiabatic invariant. It is interesting to
investigate how extremal black holes would be quantized. Discrete
spectra arise in quantum mechanics in the presence of a
periodicity  in the classical system Which in turn leads to the
existence of an adiabatic invariant  or action variable.
Boher-Somerfeld quantization implies that this adiabatic invariant
has an equally spaced spectrum in the semi-classical limit. In
this letter we have considered the extremal RN black hole in four
dimensional spacetime, using the results  for highly damped
quasi-normal modes, we obtained the area and entropy spectrum of
event horizon. Here we accepts the proposed correspondence between
the quasi-normal modes frequencies and a transition energy $\Delta
M$, we have finds that the quantum area should be $\Delta A=\hbar
\ln3$. Although the real parts of highly damped quasi-normal modes
for schwarzschild and extremal RN black hole is equal \cite{and}
$\omega_{R}=\frac{\ln3}{8\pi M }$ as one can see for example in
\cite{{kun},{bir},{and}} $\Delta A=4\hbar \ln3$ for schwarzschild
black hole. Therefore in contrast with claim of \cite {and}$
\Delta A $ is not universal for all black holes. Also Abdalla et
al \cite {ab} have been shown that the results for spacing of the
area spectrum for near extreme Kerr and near extreme
schwarzschild- de Sitter black holes differ from that for
schwarzschild, as well as for non-extreme Kerr black holes.
Although such a difference for problem under consideration in
\cite{ab} as the authors have been mentioned may be justified due
to the quite different nature of the asymptotic quasi-normal mode
spectrum of the near extreme black hole, in our problem the real
parts of highly damped quasi-normal modes for schwarzschild and
extremal RN black hole is equal. According to Eq.(\ref{ext})the
location of horizon for extreme RN black hole is in $r=M$, but
schwarzschild black hole horizon located in $r=2M$, therefore the
factor 4 in area quantum of schwarzschild black hole $\Delta
A=4\hbar \ln3$, come from the factor 2 in $r=2M$.\\
In the other hand if we assume that $\Delta A$ is indeed universal
\cite{and} and thus remains as in the Schwarzschild case $\Delta
A=4\hbar \ln3$, then the real part of the quasinormal frequency
for extremal RN black hole is as $\omega_{R}^{RN}=\frac{\ln 3}{\pi
R_{H}}$, which is different from  schwarzschild black hole case.
  

\vspace{3mm}




  \vspace{3mm}








\begin{thebibliography}{99}
\bibitem{bek1} J. D. Bekenstein,  Lett. Nuovo Cimento {\bf
11}, 467 (1974).
\bibitem{bek2} J. D. Bekenstein and V. F. Mukahnov,
Phys. Lett. {\bf B360}, 7 (1995).
\bibitem{kastrup} H. A. Kastrup,  Phys. Lett. {\bf B385}, 75 (1996).
\bibitem{bekenstescola}
J. D. Bekenstein, gr-qc/9808028.
\bibitem{hod} S. Hod,  Phys. Rev. Lett. {\bf 81}, 4293 (1998).
\bibitem{dry} O. Dreyer, Phys. Rev. Lett. {\bf 90}, 081301
(2003).
\bibitem{kun} G. Kunstatter, Phys. Rev. lett. {\bf 90}, 161301,
(2003).
\bibitem{motl} L. Motl,  gr-qc/0212096.
\bibitem{cor} A. Corichi, gr-qc/0212126.
\bibitem{ab}E. Abdalla, K. H. C. Castello-Branco, A. Lima-Santos,
gr-qc/0301130.
\bibitem{Cardoso} V. Cardoso and J.P.S. Lemos, Phys. Rev. {\bf D67}, 084020,
(2003).
\bibitem{Hod2} S. Hod, Phys. Rev. {\bf D67}, 081501,  (2003).
\bibitem{Berti}  E. Berti and K.D. Kokkotas, hep-th/0303029.
\bibitem{Polychronakos} A.P. Polychronakos, hep-th/0304135.
\bibitem{Cardoso2} V. Cardoso, R. Konoplya,  and J.P.S. Lemos, gr-qc/0305037.
\bibitem{bir}D. Birmingham, hep-th/0306004.
\bibitem{chandra}
S. Chandrasekhar, {\it The Mathematical Theory of Black Holes},
Cambridge University Press (1983).
\bibitem{kokkotas}
K. D. Kokkotas e B. G. Schmidt,  Living Reviews in Relativity
(1999);\\
H.-P. Nollert,  Class. Quant. Grav. {\bf 16}, R159-R216 (1999).
\bibitem{reggeW} T. Regge, J. A. Wheeler, Phys. Rev. {\bf 108},
1063 (1957).

\bibitem{Chandra1} S. Chandrasekhar, and S. Detweiler,
 Proc. R. Soc. London, Ser. A {\bf 344}, 441 (1975).
\bibitem{Horowitz-Habeny} G. T. Horowitz and V. Hubeny, Phys. Rev {\bf D62} 024027
(2000).
\bibitem{CFT-Correspond} V. Cardoso and J. P. S. Lemos Phys. Rev. {\bf D63} 124015
(2001).
\bibitem{ch} J. S. F. Chan and R. B. Mann, Phys. Rev. {\bf D59} 064025
(1999).
\bibitem{bir1} D. Birmingham, I. Sachs, and S. N. Solodukhin, Phys.Rev. Lett {\bf88} 151301
(2002).
\bibitem{bir2} D. Birmingham, I. Sachs, and S. N. Solodukhin,
PRDinpress hep-th/0212308 (2002).
\bibitem{ko1} R. A. Konoplya,
Phys. Rev. {\bf D66} 044009 (2002).
 \bibitem{ko2} R. A. Konoplya, Phys. Rev.{\bf D66}
084007 (2002).
 \bibitem{star} A. O. Starinets, Phys. Rev. {\bf D66} 124013
(2002).
\bibitem{ar} R. Aros, C. Martinez, R. Troncoso, and J. Zanelli,
Phys. Rev. {\bf D67} 044014 (2003).
 \bibitem{fer} S. Fernando, hep-th/0306214.

\bibitem{Moss-Norman} I. G. Moss and J. P. Norman, Class. Quant. Grav. {\bf19} 2323 (2002)
\bibitem{BHCC} R.A.Konoplya, Phys. Lett. {\bf B550} 117 (2002)
\bibitem{kim} W. T. Kim and J. J. Oh, Phys. Lett. {\bf B514} 155 (2001)
\bibitem{bek3} J.D. Bekenstein, gr-qc/9710076.
\bibitem{nitz}A. Neitzke, hep-th/0304080.
\bibitem{and} N. Andersson, C. J. Howls, gr-qc/0307020.
\end{thebibliography}
\end{document}